\documentclass[10pt,a4paper]{article}
\usepackage{opex3}
\usepackage{amsmath,amssymb}
\usepackage{amsfonts}
\usepackage{bm}
\usepackage[utf8]{inputenc}

\newcommand{\abs}[1]{\left|#1\right|}

\renewcommand{\exp}[1]{e^{#1}}

\renewcommand{\arg}[1]{\text{arg}\left(#1\right)}

\newcommand{\mat}[4]{\left[\begin{array}{cc} #1 & #2 \\ #3 &  #4  \end{array}\right]}

\begin{document}
\title{Programmable multiport optical circuits in opaque scattering materials.}

\author{Simon R. Huisman,$^{1,2}$ Thomas J. Huisman,$^{1,4}$ \\ Tom A. W. Wolterink,$^{1,3}$ Allard P. Mosk,$^{1}$ and Pepijn W. H. Pinkse$^{1, \ast}$}
\address{
$^{1}$Complex Photonic Systems (COPS), MESA$^+$ Institute for Nanotechnology,
University of Twente, P.O. Box 217,
7500 AE Enschede, The Netherlands. \\

$^{2}$Optical Sciences (OS), MESA$^+$ Institute for Nanotechnology,
University of Twente, P.O. Box 217,
7500 AE Enschede, The Netherlands. \\

$^{3}$Laser Physics and Nonlinear Optics (LPNO), MESA$^+$ Institute for Nanotechnology,
University of Twente, P.O. Box 217,
7500 AE Enschede, The Netherlands. \\

$^{4}$Institute for Molecules and Materials, Radboud University Nijmegen, Heyendaalseweg 135, 6525 AJ Nijmegen, The Netherlands.}
\email{$^{\ast}$p.w.h.pinkse@utwente.nl} 
\homepage{www.adaptivequantumoptics.org}


\begin{abstract*}
We propose and experimentally verify a method to program the effective transmission matrix of general multiport linear optical circuits in random multiple-scattering materials by phase modulation of incident wavefronts. We demonstrate the power of our method by programming linear optical circuits in white paint layers with 2 inputs and 2 outputs, and 2 inputs and 3 outputs. Using interferometric techniques we verify our ability to program any desired phase relation between the outputs. The method works in a deterministic manner and can be directly applied to existing wavefront-shaping setups without the need of measuring a transmission matrix or to rely on sensitive interference measurements.
\end{abstract*}



\section{Introduction.}

In many optical experiments light propagates through linear optical circuits, such as waveguides and interferometers. These optical circuits are often realized either in $(i)$ free-space setups containing, \textit{e.g.}, mirrors, lenses, polarizers, wave plates, or in state-of-the-art $(ii)$ integrated photonics, such as coupled waveguides and cavities \cite{Hecht2002, Saleh2007}. Both free-space and integrated optical circuits are robust platforms for performing experiments with low optical losses. In principle, arbitrary complex linear circuits can be built this way \cite{Beck1994, Miller2013}. However, once the optical circuit has been built, one has often little flexibility in modifying or programming the functionality, especially in a running experiment. One can design the experiment to partly circumvent this issue by including adaptive optical elements, which mostly give a controllable (phase) delay. Especially in integrated photonics much effort is invested in controlling the refractive index by, \textit{e.g.}, temperature tuning, free-carrier excitation, or optical Kerr switching \cite{Leonard2002, Zhang2011, Shadbolt2012, Bonneau2012, Yuce2012}. Nevertheless, it becomes a major challenge if one would like to give the optical circuit a functionality that is entirely different from its original design, \textit{e.g.}, changing the number of input and output modes and the correlations between them.

\begin{figure}[]
    \includegraphics[width=12.5 cm]{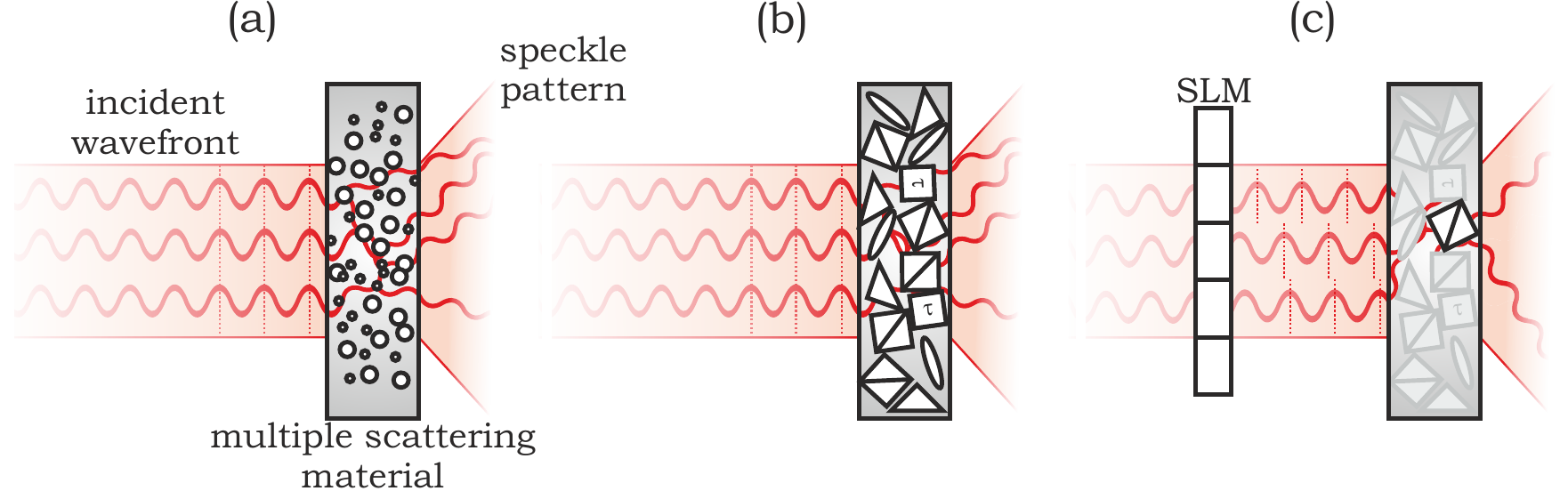}
\caption{\textbf{Wavefront-shaped programmable linear optical circuits.} $(a)$ Incident light on a multiple-scattering medium results in a speckle pattern. $(b)$ The scattered light can be described by a scattering matrix, representing a complicated linear optical circuit. The scattering matrix is here represented as light propagating through an effective medium with the same correlations as the optical circuit, however, these optical elements are not physically located at these positions in the material. $(c)$ By phase modulation of the incident wavefront with a spatial light modulator (SLM) it becomes possible to address correlations in the scattering matrix to create an interference pattern with a desired functionality. In this picture light travels through the material as if it would have traveled through a beam splitter. Note: reflection is omitted in this figure for clarity.}
\label{Fig_1}
\end{figure}

Here we suggest the radical different approach to use wavefront shaping of light on random multiple-scattering materials as a platform for programmable linear optical circuits, as is illustrated in Fig. \ref{Fig_1}. Incident coherent light on stationary random multiple-scattering media, such as white paint, teeth, and paper, gives rise to speckle patterns as the result of the collective interference of scattered waves. The individual far-field speckle spots form diffraction-limited beams that are correlated to each other as if light would have propagated through a very complicated random linear optical circuit \cite{Freund1990}. Wavefront shaping is an adaptive optical technique in which an incident wavefront on a scattering medium is spatially modulated to intensity enhance one or multiple speckle spots \cite{Vellekoop2007, Mosk2012}. In essence one controls by modulating, \textit{e.g.}, the phase of the incident wavefront the degree of mode mixing of all scattered waves that contribute constructively to the target spots. Wavefront shaping has been generally known for focusing and imaging with multiple-scattering media. Wavefront shaping has also transformed the speckle patterns of multiple-scattering materials to behave like many linear optical components, such as waveguides and lenses \cite{Vellekoop2007, Putten2011}, optical pulse compressors \cite{Aulbach2011, Katz2011}, programmable wave plates \cite{Guan2012}, and recently also beam splitters \cite{Huisman2014}. The manner in which the incident input modes are projected to the output modes by the scattering material is described in the scattering matrix $\textbf{S}$. By wavefront shaping one can address subsets of the scattering matrix to project on desired modes to obtain the functionality of the desired linear optical device. Since one is in general not able to control all incident modes of the scattering matrix, and the scattering matrix might not contain all desired correlations, one has to tolerate losses that are typically higher than of custom fabricated optical circuits. On the other hand, this optical circuit is inherently programmable in functionality, has a system size comparable to integrated photonics, and is in terms of optical hardware very easy to implement and adapt. In contrast to most (integrated) multi-mode interference based devices \cite{Soldano1995}, our method exploits disorder for functionality and is therefore robust against imperfections and does not require careful fabrication of the scattering structure. However, up to now, a general method for creating arbitrary multiport linear optical circuits by wavefront shaping in random scattering materials was still missing. A promising strategy is to measure the transmission matrix of the sample \cite{Popoff2010} and to adapt the wavefront to program the desired interference pattern. Unfortunately this requires many interference measurements, which is not always feasible.

We present a general wavefront-shaping method that controls the phase and amplitude correlations between the enhanced target spots. For this method it is not necessary to measure a transmission matrix or to rely on sensitive interference measurements. One only requires a phase-only spatial light modulator (SLM) and a camera; both are already present in most wavefront-shaping setups. We program in a deterministic manner an interference pattern that represents the functionality of multiport linear optical circuits, where the light interferes in a compact system size comparable with integrated photonics. We demonstrate the power of our method by wavefront shaping equivalents of $2\times 2$ and $2\times 3$ linear optical circuits using a layer of strongly scattering white paint deposited on glass. Our input basis consists of wavefront-shaped beams and the output basis consists of individual spots.

\begin{figure}[]
  \includegraphics[]{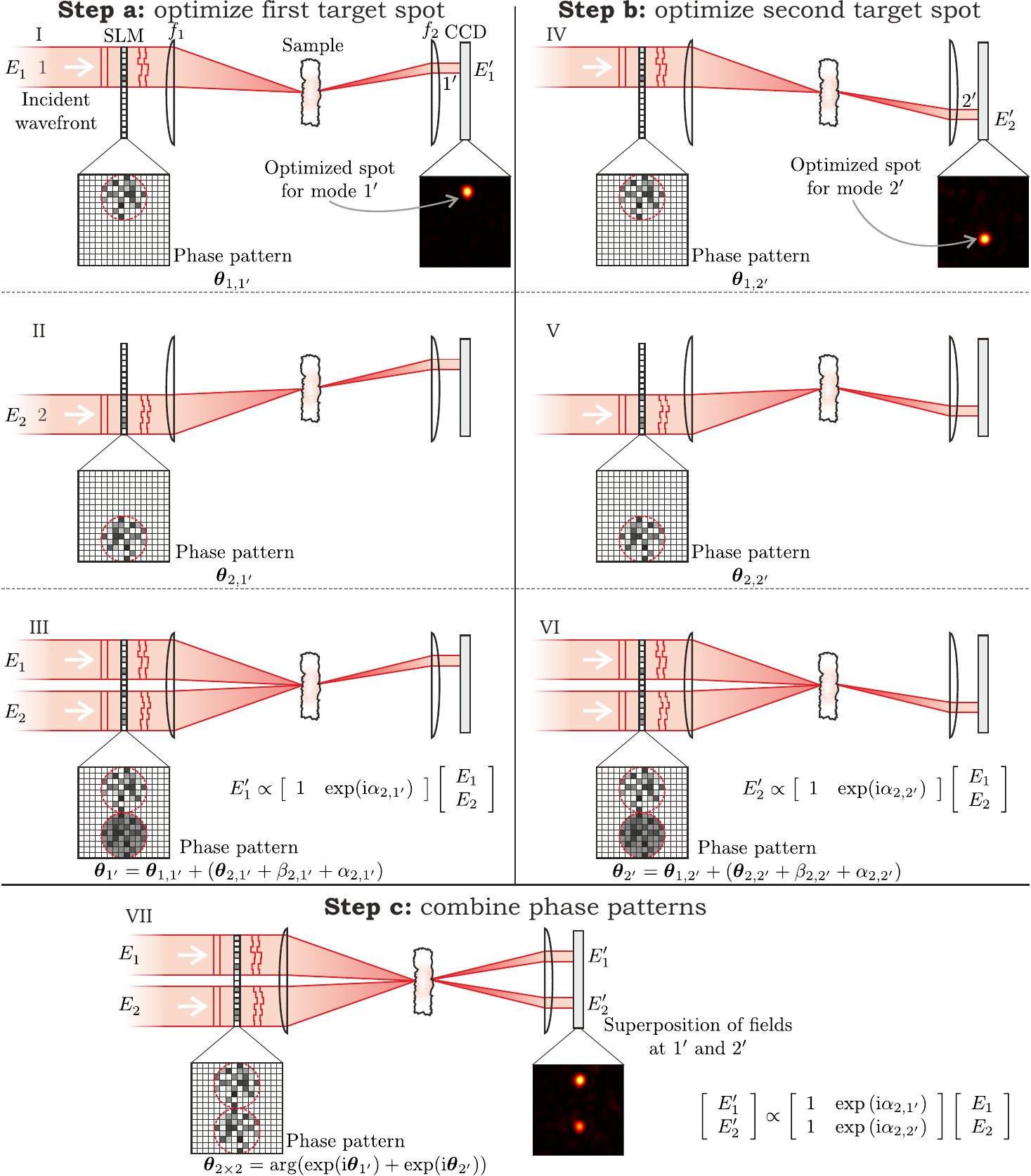}
\caption{\textbf{Schematic illustration for programming a $2\times 2$ linear optical circuit.} The incident input modes $1$ and $2$ are spatially separated on the SLM. (I-III) Optimization for output mode $1^{\prime}$ providing phase pattern $\boldsymbol{\theta}_{1^{\prime}}$. (IV-VI) Optimization for output mode $2^{\prime}$ providing phase pattern $\boldsymbol{\theta}_{2^{\prime}}$. (VII) Finally one writes phase pattern $\boldsymbol{\theta}_{2 \times 2}=\arg{{\exp{{\rm{i} \boldsymbol{\theta}_{1^{\prime}}}}+\exp{{\rm{i} \boldsymbol{\theta}_{2^{\prime}}}}}}$ to obtain a superposition of the fields in steps I and II. The CCD pictures are snapshots of our experiments on the $2\times 2$ optical circuit.}
\label{Fig_2}
\end{figure}

\section{Algorithm.}

We propose a method that is based on linear scattering of light in random multiple-scattering materials. In the following we introduce our procedure, or algorithm, that describes the consequent steps to be performed on the setup to arrive at the desired circuits. We describe here the most general implementation of our algorithm that should be valid for any existing wavefront-shaping setup. The algorithm is illustrated for a $2 \times 2$ optical circuit in Fig. \ref{Fig_2} and explained for the general $n \times m$ optical system with $n$ separate inputs and $m$ separate outputs. We assume for simplicity that a single phase-only spatial light modulator is used, the incident input modes are spatially separated on the same SLM surface, and the resulting interference pattern is observed with a CCD camera. We use the term 'optimization' of a spot for intensity enhancing a target spot in a speckle pattern by phase modulation of the incident light. A single spot is considered as one independent output mode: the spots form an orthogonal basis. A single incident wavefront shaped beam is considered to be a single input mode of the system. We assume that the desired optical circuit is supported by the scattering matrix of the multiple-scattering materials. The algorithm consists of the following steps:
\begin{enumerate}
\item Start with the first input mode incident on the SLM and the sample. Optimize by phase modulation of the incident light a target spot that forms output mode $1^\prime$. Examples of algorithms for optimizing a single spot are described in Refs. \cite{Vellekoop2007, Vellekoop2008-2, Yilmaz2013}. Store the corresponding phase pattern on the SLM as $\boldsymbol{\theta}_{1,1^\prime}$. This corresponds with Fig. \ref{Fig_2} step I.

\item Repeat step 1 to optimize an enhanced spot $1^\prime$ for each other input mode. Store the final phase patterns $\boldsymbol{\theta}_{2,1^\prime} \cdots \boldsymbol{\theta}_{n,1^\prime}$. This corresponds with Fig. \ref{Fig_2} step II. Note: the phase pattern should only modulate the corresponding input mode, \textit{e.g.}, $\boldsymbol{\theta}_{2,1^\prime}$ acts on input mode $2$ and not on input mode $1$ since this mode is located at a different location on the SLM. At the location of input mode 1, $\boldsymbol{\theta}_{2,1^\prime}=0$, etc.

\item Create a phase pattern:
\begin{equation}
\boldsymbol{\theta}_{1,1^\prime}+\boldsymbol{\theta}_{i,1^\prime} + \beta_{i,1^\prime},
\label{PhasePatternExpr}
\end{equation}
with $i = 2$. Now we have input mode $1$ and input mode $2$ incident on the SLM and sample, all others input modes are blocked. Add a phase offset $\beta_{i,1^\prime}$ at the corresponding illuminated pattern of the second input mode on the SLM (gray filled circle in Fig. \ref{Fig_2} step III) that maximizes the intensity in output spot $1^\prime$. The intensity is maximal when mode 1 and mode 2 are projected in phase on $1^\prime$. Store this value for phase $\beta_{i,1^\prime}$.

\item Now block input mode 2 and open input mode 3, and repeat the procedure of the previous step for mode 1 and mode 3. Redo this for all remaining input modes. For $n\ge2$ expression \ref{PhasePatternExpr} can be generalized:
\begin{equation}
\boldsymbol{\theta}_{1^\prime}=\sum_{i=2}^{n}(\boldsymbol{\theta}_{i,1^\prime} + \beta_{i,1^\prime})+\boldsymbol{\theta}_{1,1^\prime}.
\end{equation}

\item In steps 3-4 we have enforced that all input modes are projected in phase to the target spot, where the first input mode was used as reference. Next we have all input modes incident on the sample with phase mask $\boldsymbol{\theta}_{1^\prime}$ on the SLM. In principle all input modes can be optimized simultaneous per output mode as long as the input modes are interferometric stable during optimization and that this automatically projects all input modes in phase to the target spot. The corresponding complex field amplitude $E^\prime_i$ at the target spot $1^\prime$ becomes:

\begin{equation}
E^\prime_i=\exp{\rm{i}\phi_{1^\prime}} \left[\begin{array}{cccc} \abs{t_{1,1^\prime}} & \abs{t_{2,1^\prime}} & \ldots & \abs{t_{n,1^\prime}} \end{array}\right] \left[\begin{array}{c} E_1\\ E_2 \\ \vdots \\ E_n \end{array}\right],
\label{B1}
\end{equation}

where $\phi_{1^\prime}$ is an overall phase factor with respect to a fixed reference and we have ignored an overall normalization factor. The amplitudes $\abs{t_{i,1^\prime}}$ should be approximately equal to each other for an isotropic random scattering material, and is given by the square root of the intensity of the optimized spot. This equation can be simplified to a transfer matrix equation describing a $n \times 1$ optical system:

\begin{equation}
E^\prime_i=\textbf{T}_{n\times 1,1^\prime} \left[\begin{array}{c} E_1\\ E_2 \\ \vdots \\ E_n \end{array}\right].
\end{equation}

\item \textit{Amplitude control of each element in $\textbf{T}_{n\times 1,1^\prime}$}: One can achieve amplitude control by manipulating the intensity enhancement of the spot for each input mode. Suppose one wants to decrease $\abs{t_{2,1^\prime}}$. One adds to phase pattern $\boldsymbol{\theta}_{1^\prime}$ at the location of the second input mode a random phase pattern with a controlled amplitude to reduce the intensity enhancement to the desired level. One stores this new phase pattern as $\boldsymbol{\theta}_{1^\prime}$. The phase of the transfer matrix element should remain unaffected if there are sufficiently SLM segments used for the mode (about $\sim 10^2$ segments). Otherwise one can compensate for this additional phase shift by repeating the procedure of step 4 for the specific input mode. It is important to note that this manner of controlling the amplitude of $\abs{t_{i,1^\prime}}$ will only reduce the amplitude level.

\item \textit{Phase control of each element in $\textbf{T}_{n\times 1,1^\prime}$}: One can achieve phase control over each input mode $i$ by writing a desired phase offset $\alpha_{i,1^\prime}$ to  $\boldsymbol{\theta}_{1^\prime}$ at the corresponding illuminated region on the SLM. In this manner the unnormalized transfer matrix $\textbf{T}_{n\times 1,1^\prime}$ becomes:

\begin{equation}
\textbf{T}_{n\times 1,1^\prime}=\exp{\rm{i}\phi_{1^\prime}} \left[\begin{array}{cccc} \abs{t_{1,1^\prime}} \exp{\rm{i}\alpha_{1,1^\prime}} & \abs{t_{2,1^\prime}} \exp{\rm{i}\alpha_{2,1^\prime}} & \ldots & \abs{t_{n,1^\prime}} \exp{\rm{i}\alpha_{n,1^\prime}} \end{array}\right].
\label{B4}
\end{equation}
We have now controlled independently both the phase and amplitude of each element in $\textbf{T}_{n\times 1,1^\prime}$. This is also illustrated in Fig. \ref{Fig_2} step III. In the remainder of this section we will explain how to get a desired transmission matrix with multiple inputs and multiple outputs.

\end{enumerate}

\begin{enumerate}
\setcounter{enumi}{7}

\item Repeat steps 1-7 for the remaining number of orthogonal output spots. This is also illustrated in Fig. \ref{Fig_2} steps IV-VI. At the end of this step you have $m$ independent transfer matrices $\textbf{T}_{n\times 1,m^\prime}$ that each describe an $n \times 1$ optical system.

\item With all input modes incident, write the phase pattern $\boldsymbol{\theta}_{n \times m}$:

\begin{equation}
\boldsymbol{\theta}_{n \times m}=\arg{\sum_{j=1}^m \boldsymbol{{c_j}} \exp{\rm{i} \boldsymbol{\theta}_\textit{j}}},
\end{equation}
where subscript $j$ is a label that is $1$ for the first output mode, with $j \leq m$. The field in target output modes become related to the input modes as:
\begin{equation}
\left[\begin{array}{c} E^\prime_1\\ E^\prime_2 \\ \vdots \\ E^\prime_m \end{array}\right] = f_1
\left[\begin{array}{c} c_1 \textbf{T}_{n\times 1,1^\prime} \\ c_2 \textbf{T}_{n\times 1,2^\prime}  \\ \vdots \\ c_m \textbf{T}_{n\times 1,m^\prime}  \end{array}\right]
\left[\begin{array}{c} E_1\\ E_2 \\ \vdots \\ E_n \end{array}\right],
\end{equation}
with $f_1$ a normalization factor. In this manner we have programmed an unnormalized transmission matrix $\bold{T}_{m\times n}$ given by:
\begin{equation}
\bold{T}_{m\times n}=\left[\begin{array}{c} c_1 \textbf{T}_{n\times 1,1^\prime} \\ c_2 \textbf{T}_{n\times 1,2^\prime}  \\ \vdots \\ c_m \textbf{T}_{n\times 1,m^\prime}  \end{array}\right].
\end{equation}
This is also illustrated for the $2 \times 2$ system in Fig. \ref{Fig_2} VII.
\end{enumerate}

In Eq. (5) the superposition of the individual incident field patterns results into the desired correlations between the output modes because the scattering material is a linear system. This procedure requires that the output spots behave as independent uncorrelated output modes. If this is the case, all individual matrices describing patterns $\exp{\rm{i} \boldsymbol{\theta}_\textit{j}}$ in Eq. (5) should be orthogonal to each other. This condition is met if all $\exp{\rm{i} \boldsymbol{\theta}_{\textit{i},\textit{j}^\prime}}$ for a given input mode $i$ are orthogonal to each other. In practice this condition is not met since the spots are weakly correlated with each other, which can be caused by the correlations in the scattering matrix of the material itself \cite{Sheng1995, Akkermans2007}, or by the fact that one only addresses subsets of the scattering matrix of the sample by wavefront shaping \cite{Vellekoop2008}. However, if this subset is large enough (typically $\sim 10^2$ independent channels per wavefront describing an input mode), the spots behave approximately orthogonal to each other and therefore also the phase masks become orthogonal and the algorithm will work.

\begin{figure}[]
  \includegraphics[]{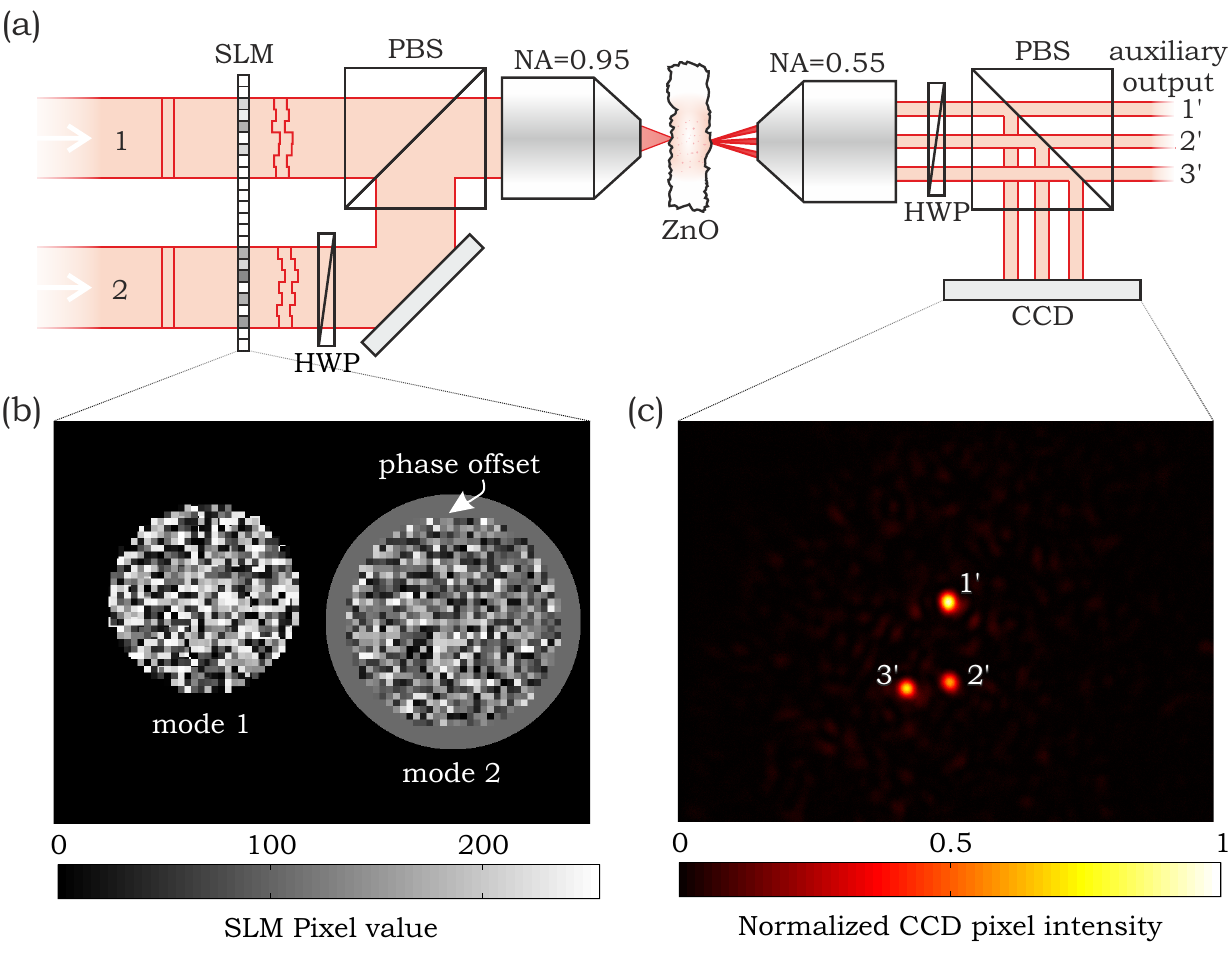}
\caption{\textbf{Setup for wavefront-shaped optical circuits.} $(a)$ Two input modes $(1,2)$ are phase-modulated with a spatial light modulator (SLM). Both modes are spatially overlapped with a polarizing beam-splitter cube (PBS). The modes are focused on a layer of white paint (ZnO particles) that has been spray coated on a 1.5 mm thick microscope slide. The transmitted light is projected on a CCD camera. Three output modes $1^{\prime}$, $2^{\prime}$, and $3^{\prime}$ are selected. $(b)$ Optimized phase pattern on the SLM. A phase offset is applied to the second incident mode. $(c)$ Camera image for three optimized spots when both input modes are incident on the phase pattern of $(b)$.}
\label{Fig_3}
\end{figure}

\section{Experimental setup and methods.}

We demonstrate our method by wavefront shaping a programmable $2\times2$ and $2\times 3$ transmission matrix in a layer of white paint. In this section we describe our experimental setup and our measurement method.

\subsection{Experimental setup.}

Our setup is illustrated in figure \ref{Fig_3}$(a)$. This is the same setup as described in Ref. \cite{Huisman2014}. The light source is a mode-locked Ti:Sapphire laser (Spectra-Physics, Tsunami) emitting transform-limited pulses at a repetition rate of $80$ MHz with a pulse width of approximately $0.3$ ps and a center wavelength of $790.0$ nm. The beam is split and coupled into two separate single-mode fibers. The output of the two fibers have identical polarization and beam waist and form the input modes $1$ and $2$. The two modes are phase-modulated with a SLM (Hamamatsu, LCOS-SLM). Figure \ref{Fig_3}$(b)$ illustrates an optimized phase pattern. Clearly, both modes are separated from each other on the SLM surface. After leaving the SLM, the two modes are spatially overlapped with a half-wave plate (HWP) and polarizing beam splitter (PBS) cube, resulting in collinear propagation of modes with orthogonal polarization. This allows us to completely fill the aperture of the objective (NA=0.95) that images the SLM on the conjugate plane of the layer of white paint. The surface of the SLM is imaged on the back focal plane of the objective with two lenses in a 4-focal-length-configuration (not shown). Both pulses arrive simultaneously at the sample to within $20$ fs. We make sure that the power of both input modes on the objective are identical (approximately $0.5$ mW per mode). However, input mode 1 is transmitted more efficiently by the objective than input mode 2 because of experimental imperfections. This causes the optimized spots for mode 1 to have a higher intensity. The layer of white paint consists of ZnO powder with a scattering mean free path of $0.7 \pm 0.2$ $\mu$m. The layer is approximately $30$ $\mu$m thick and spray painted on a glass microscope slide of 1.5 mm thickness. The transmitted pattern is collected with a second objective (NA=0.55) and directly projected on a CCD camera after reflection on a PBS, see for example Fig. \ref{Fig_3}$(c)$ where three optimized spots are visible. The intensity values for the CCD pixels that correspond to the target spots are spatially integrated to obtain the output powers for the interference pattern. The optimized spots can be transmitted through the PBS, towards a different part of the setup for applications, by rotating the HWP.

\subsection{Measurement method.}

We start with a single input mode incident on the material and selected on the camera a location for intensity enhancing a target spot. The SLM is divided into segments of 10x10 pixels. One input mode is controlled by approximately $500$ segments (dividing the SLM into smaller segments did not significantly improve the intensity enhancement). The SLM controls the phase by addressing the pixels with 8-bit pixel values.  Therefore all phase results are presented in pixel values and the corresponding value in radians. In our experiment a pixel value of 207 corresponds with a phase difference of $2 \pi$ rad. We first pre-optimize spots by fitting the optimal phase for each segment that provides maximum constructive interference in the target spot, as used in Ref. \cite{Vellekoop2007}. We apply this method twice, in spirit of the work in Ref. \cite{Yilmaz2013}. Afterwards a final optimization is made by sequentially addressing each segment with a random phase. This phase value is accepted if the intensity increased. We apply this procedure about 5 times for each pixel. In case there is no intensity present at the intended target, the procedure is also used as an initial optimization. The total optimization time for one input mode to one output mode is approximately 1.5 hours. The typical intensity enhancement for an individual spot is in the order of $50$ times compared to the average intensity of the other spots.

\begin{figure}[]
  \includegraphics[]{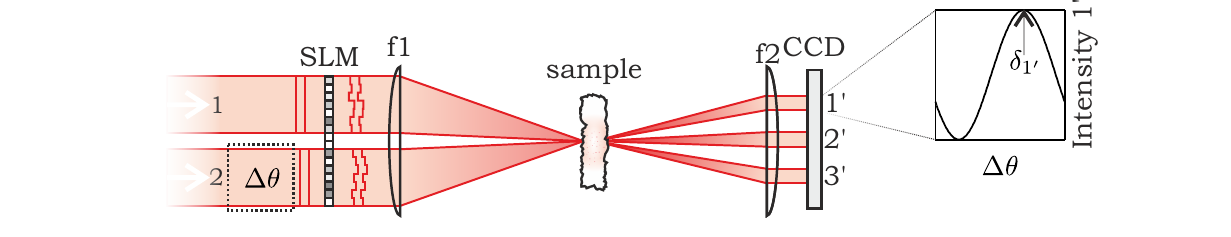}
\caption{\textbf{Interference measurement of the effective transmission matrix.} A phase difference $\Delta \theta$ is applied between the two input modes. The intensity in the target optimized spots is measured as a function of this phase difference.}
\label{Fig_4}
\end{figure}

This procedure is repeated for each input mode and each output mode and with the algorithm the desired transmission matrix was constructed. To confirm that the input modes and the optimized spots are correlated as programmed in the transmission matrix, we perform interference measurements, similar as described in Ref. \cite{Huisman2014}, which is illustrated in Fig. \ref{Fig_4}. A phase difference $\Delta \theta$ is applied between the input modes to monitor the intensity and resulting interference in the output modes. From these interference measurements one can extract in principle the phase relation between the output modes and the transmission amplitude coefficients. This phase difference $\Delta \theta$ is applied with the SLM on input mode 2. Figure \ref{Fig_3}$(b)$ shows an example. We give this phase offset a bigger area on the SLM to avoid edge effects. We define $\delta_{j^\prime}$ the phase $\Delta \theta$ for which maximum intensity occurs in output mode ${j^\prime}$, as indicated for mode $1^\prime$ in Fig. \ref{Fig_4}$(b)$. In our experiments we only extract the phase relation between output modes.

\section{The $2\times2$ optical circuit.}
\begin{figure}[]
  \includegraphics[]{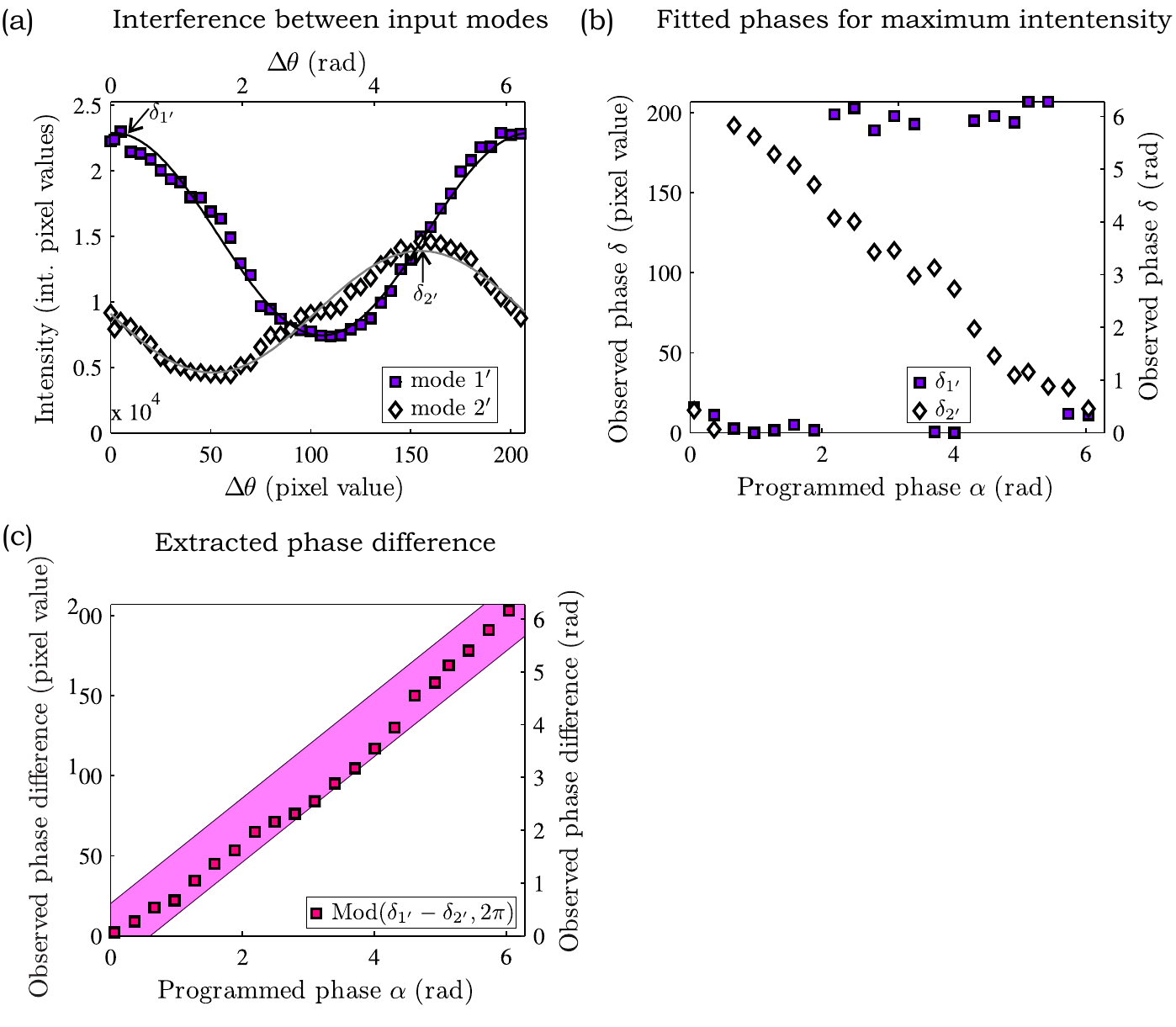}
\caption{\textbf{Experimental realization of a programmed $2 \times 2$ linear optical circuit.} A transmission matrix is programmed for which output mode $2^\prime$ has a programmable phase difference $\alpha$ with respect to output mode $1^\prime$. $(a)$ Example of a measured (symbols) interference characterization of the transmission matrix for $\alpha = 2 $ rad. Sine fits (solid) are used to determine the phases $\delta$ for which maximum intensity occurs. $(b)$ Extracted phases $\delta$ as a function of the programmed phase $\alpha$. $(c)$ Extracted phase difference between the output modes (symbols) in comparison with the expected phase difference (diagonal band) based on the programmed phase $\alpha$. The observed phase differences between the output modes match the programmed phase differences excellently.}
\label{Fig_5}
\end{figure}

\begin{figure}[]
  \includegraphics[]{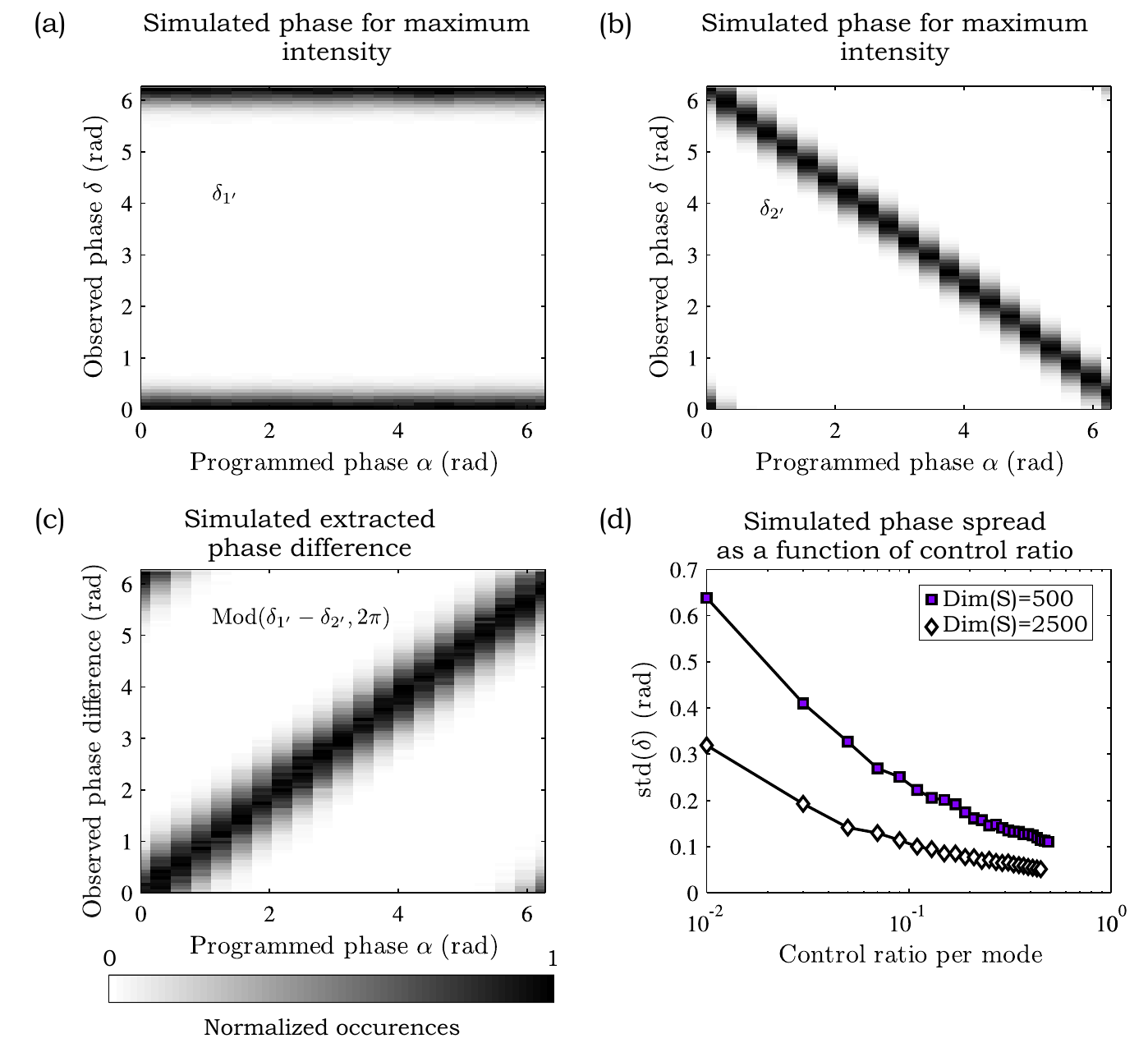}
\caption{\textbf{Computational results on the phase differences of the $2 \times 2$ linear optical circuit.} 10,000 Realizations were simulated for systems with a scattering matrix of dimension 1000 and with 50 controlled channels per input mode. $(a)$ Obtained distribution for phase $\delta_{1^\prime}$. $(b)$ Obtained distribution for phase $\delta_{2^\prime}$. $(c)$ Extracted phase difference between the output modes. $(d)$ Width of the phase distributions as a function of controlled channels for scattering matrices of dimension 500 and 2500.}
\label{Fig_6}
\end{figure}

We first demonstrate our algorithm by programming a $2 \times 2$ optical circuit with a transmission matrix of the form:
\begin{equation}
\bold{T}_{2\times2}= \mat{\abs{T_{11}}}{\abs{T_{12}}}{\abs{T_{21}}}{\abs{T_{22}}\exp{\rm{i} \alpha}}.
\label{2x2}
\end{equation}
For simplicity we only control the phase correlations inside the transmission matrix and not the amplitude. The amplitudes $\abs{T_{a,b}}$ are set by the intensity enhancement of the individual spots and are approximately equal. We let the phase of element $\bold{T}_{2\times2}(2,2)$ vary with controllable phase $\alpha$, to demonstrate the ability to program the phase correlations of this optical circuit. The four phase patterns $\boldsymbol{\theta}_{1,1^\prime}$, $\boldsymbol{\theta}_{2,1^\prime}$, $\boldsymbol{\theta}_{1,2^\prime}$, $\boldsymbol{\theta}_{2,2^\prime}$ were combined to overall phase pattern $\boldsymbol{\theta}_{2\times2}(\alpha)$  following the procedure of our algorithm, in which step 6 was excluded. Phase $\alpha$ is varied in 21 steps ranging from pixel value 0 to 210 with steps of 10, so therefore in total 21 different phase patterns $\boldsymbol{\theta}_{2\times2}(\alpha)$ were made for our measurements. Note that the transmission matrix for $\alpha=\pi$ represents the equivalent of a standard optical beam splitter \cite{Campos1989}, which is in this case fully controlled, in contrast to the algorithm for the wavefront-shaped beam splitters obtained in Ref. \cite{Huisman2014}.

The results are presented in Fig. \ref{Fig_5}. Figure \ref{Fig_5}$(a)$ shows the interference results for the two optimized spots for $\alpha = 2$ rad. Two sine functions are fitted to the output modes to determine the phases $\delta_{1^\prime}$ and $\delta_{2^\prime}$ at which maximum intensity occurs. The phases $\delta_{1^\prime}$ and $\delta_{2^\prime}$ for each $\alpha$ are shown in Fig. \ref{Fig_5}$(b)$. We observe that $\delta_{1^\prime}$ is approximately constant as function of $\alpha$, while $\delta_{2^\prime}$ decreases linearly with $\alpha$. Ideally one would expect that $\delta_{1^\prime}$ remains exactly constant as function of $\alpha$. However, there is crosstalk between the modes since they are not perfectly orthogonal. Phase fluctuations due the stability of our setup can be neglected. Figure \ref{Fig_5}$(c)$ presents the main results of the $2\times2$ optical circuit: the observed phase difference between the output modes (symbols) as a function of the programmed value for\textsc{ $\alpha$}. The diagonal band represents the expected phase based on the accuracy at which we program the transmission matrix. We observe an excellent agreement between our measurements and predicted values. Our proof-of-principle experiments demonstrate full phase control of wavefront shaping $2\times 2$ optical circuits in white paint in a deterministic manner.

For comparison, we also performed simulations to support our measurements. Figure \ref{Fig_6}$(a-c)$ presents computational results in which we have repeated virtually our experiment 10,000 times on simulated random unitary scattering matrices with a dimension of 1000 and 50 controlled channels per input mode (a scattering matrix of dimension 1000 means that there are 1000 independent channels). Similar simulations are performed by us in Ref. \cite{Huisman2014}. Figure \ref{Fig_6}$(a)$ shows the histograms for the fitted phase $\delta_{1^\prime}$ for which maximum interference occurs for output mode 1 as a function of $\alpha$. Figure \ref{Fig_6}$(b)$ shows the histograms for fitted phase $\delta_{2^\prime}$. Figure \ref{Fig_6}$(c)$ shows the histogram for the phase difference between the output modes.  All three figures demonstrate histograms with a finite width. This width is a manifestation of the non-orthogonality caused by addressing a subset of the scattering matrix, as described in step 9 of the algorithm. In these simulations a subset of the scattering matrix becomes addressed that consists of 2 rows, corresponding to the output spots, and 100 columns, corresponding to the two modes that are controlled by 50 independent channels each. This subset does not consist of orthogonal rows anymore, resulting in the output spots being weakly correlated, which also occurs in our experiment. Apparently, the width of the observed phases $\delta_1$ and $\delta_2$ is independent of $\alpha$ and is identical for $\delta_1$ and $\delta_2$.

Figure \ref{Fig_6}$(d)$ shows another set of simulations where the standard deviation of the phase distributions is plotted as a function of the controlled number of independent channels per wavefront-shaped input mode divided by the dimension of the scattering matrix of the system. We call this ratio the control ratio; suppose one wavefront-shaped input mode controls 100 channels and the dimension of the scattering matrix is 1000, then the control ratio is $10\%$. We have performed these simulations for scattering matrices with dimension 500 and 2500. Each data point was obtained by ensemble averaging over 1,000 different realizations. We observe that the standard deviation of these phase distributions decreases with an increased control ratio. This figure has important consequences for choosing a material for wavefront shaping optical circuits. Suppose one wants to wavefront shape a $2 \times 2$ optical system with reduced losses by considering a sample with less channels, \textit{e.g.}, a disordered multi-mode fiber instead of a thick layer of white paint. For the multi-mode fiber, the dimension of the scattering matrix is much smaller and therefore the width of the phase distribution increases. However, the control ratio should increase as well, which reduces the width of the phase distributions. It is intriguing that both effects counteract each other. More computational and theoretical work is necessary to understand the efficiency and accuracy of wavefront-shaping optical circuits, which is beyond the scope of this article.

\begin{figure}[]
  \includegraphics[]{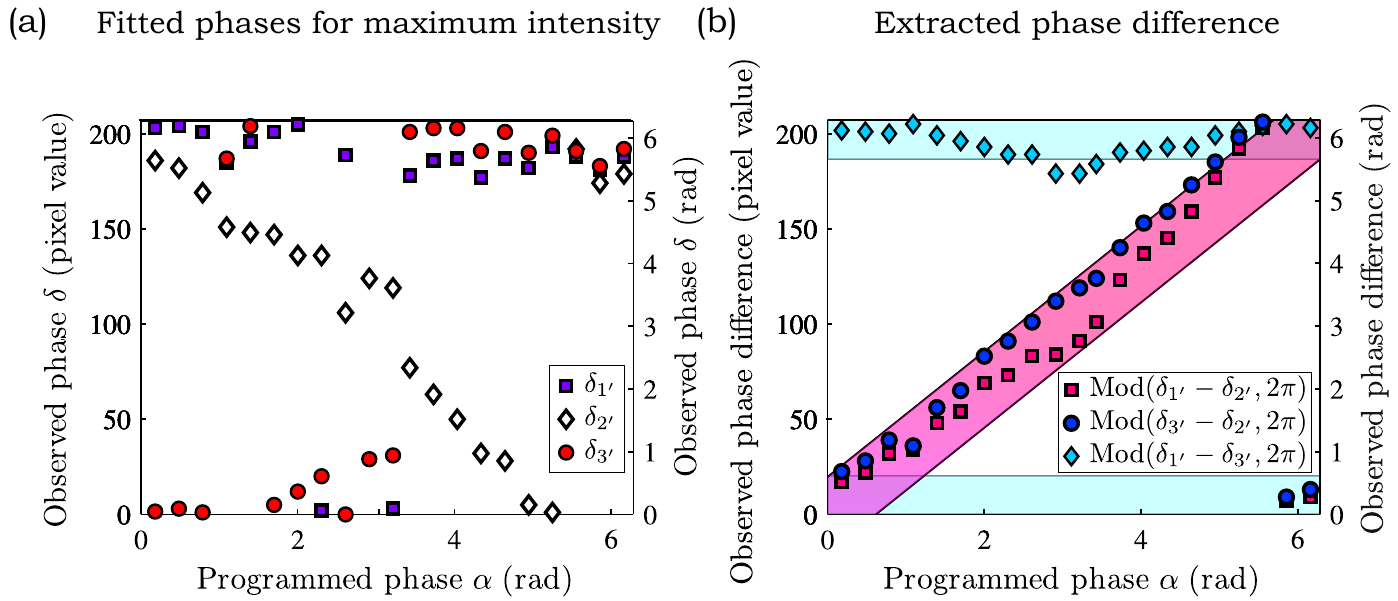}
\caption{\textbf{Experimental realization of a programmed $2 \times 3$ linear optical circuit.} A transmission matrix is programmed for which output mode $2^\prime$ has a programmable phase difference $\alpha$ with respect to output modes $1^\prime$ and $3^\prime$. $(a)$ Extracted phases $\delta$ (symbols) as a function of the programmed phase difference $\alpha$. $(b)$ Extracted phase difference between the output modes (symbols) in comparison with the expected phase differences (horizontal and diagonal bands) based on the programmed phase $\alpha$. The observed phase differences between the output modes match the programmed phase differences excellently. The pink band indicates the uncertainty in the phase determination.}
\label{Fig_7}
\end{figure}

\section{The $2\times3$ optical circuit.}
In the previous section we have presented experiments on a wavefront shaped $2 \times 2$ optical circuit. Here we present the same type of experiments for a $2\times 3$ optical circuit. The transmission matrix of the $2 \times 3$ optical circuit is given by:
\begin{equation}
\bold{T}_{2\times3}=
\left[\begin{array}{cc}
\abs{T_{11}} &  \abs{T_{12}} \\
\abs{T_{21}} &   \abs{T_{22}}\exp{\rm{i} \alpha} \\
\abs{T_{31}} & \abs{T_{32}} \\
\end{array}\right],
\label{2x3}
\end{equation}
with controllable phase difference $\alpha$. We let the phase of element $\bold{T}_{2\times3}(2,2)$ vary with controllable phase $\alpha$. Phase $\alpha$ is varied in 21 steps ranging from pixel value 0 to 210 with steps of 10.

The results are presented in Fig. \ref{Fig_7}. The phases for which maximum interference occurs in the output modes, $\delta_{1^\prime}$, $\delta_{2^\prime}$, and $\delta_{3^\prime}$, are shown in Fig. \ref{Fig_7}$(a)$ for each programmed phase $\alpha$. We observe that $\delta_{1^\prime}$ and $\delta_{3^\prime}$ are approximately constant as function of $\alpha$, while $\delta_{2^\prime}$ decreases linearly with $\alpha$. The fluctuations in $\delta_{1^\prime}$ and $\delta_{3^\prime}$ are caused by the non-orthogonality of the output modes, as was explained in more detail for the $2\times 2$ optical circuit in the previous section. Figure \ref{Fig_7}$(b)$ shows the main result, the relative phase differences between the output modes. The symbols represent the measurements and the bands represent the expected phases based on the accuracy at which we program the transmission matrix. The observed phase differences between the output modes match very well the expected phase differences. Our proof-of-principle experiments demonstrate full phase control of wavefront shaping $2\times 3$ optical circuits in white paint in a deterministic manner.

\section{Conclusions and outlook.}

In summary, we have presented a method that transforms random multiple-scattering materials into programmable multiport linear optical circuits by phase modulation of incident wavefronts. The method provides the desired transmission matrix in a deterministic manner and it can be implemented in most existing wavefront-shaping setups. We have described proof-of-principle experiments in which we have used a white paint layer as programmable $2 \times 2$ and $2\times 3$ optical circuits. The experimental observed phase relations demonstrate a very good agreement with theory.  Our method offers a very simple implementation of a programmable optical circuit which is robust against disorder and where the light interferes in a compact system size comparable with integrated photonics.

We anticipate that our method can be implemented to make more advanced linear optical circuits with a larger number of inputs and outputs. More research is required to understand how efficiently one can shape an interference pattern with a programmed correlation to achieve the functionality of the desired optical circuit. Many parameters have to be explored to identify the restrictions of our algorithm. It would be fascinating to explore the influence of the scattering properties of the material, \textit{e.g.}, the sample geometry, the sample thickness, and scattering mean free path. The performance of our algorithm is expected to be affected by the efficiency of the wavefront shaping process, which determines the intensity enhancement and the amount of light that gets focused in a target spot. Using random scattering media described by scattering matrices of lower dimension, such as disordered multi-mode fibers or planar disordered structures, will reduce optical losses caused by uncontrolled channels. This would make it possible to use these wavefront-shaped optical circuits for adaptive quantum optical experiments in multiple-scattering materials \cite{Huisman2013, Goorden2013}. In our experiments we worked with linearly-polarized light. By using polarization-selective components for the incident light and the scattered light, and additional cameras for detection, it becomes possible to use our algorithm for any polarization basis for the input and output modes, like circularly-polarized light. In addition, it would be intriguing to use structured scattering materials in order to more efficiently address certain correlations in the scattering matrix for programmed functionality.

\section*{Acknowledgments.}
We thank K. -J. Boller, S. A. Goorden, J. L. Herek, J. P. Korterik, F. B. Segerink, I. M. Vellekoop, and  W. L. Vos for discussions and support. This
work was supported by the Stichting Fundamenteel Onderzoek der Materie (FOM) that is financially supported by the Nederlandse Organisatie voor Wetenschappelijk Onderzoek (NWO). A. P. M. acknowledges ERC grant 279248. P. W. H. P. acknowledges NWO Vici.


\end{document}